\documentclass[preprint,12pt]{elsarticle}
\usepackage{amsmath}
\usepackage{amssymb}
\usepackage[pdftex,colorlinks]{hyperref}
\usepackage{multirow}

\biboptions{compress}

\begin{document}

\begin{frontmatter}

\title{Quantum election scheme based on anonymous quantum key distribution}

\author{Rui-Rui Zhou}
\author{Li Yang\corref{1}}\ead{yang@is.ac.cn}
\cortext[1]{Corresponding author.}
\address{State Key Laboratory of Information Security, Graduate University of Chinese Academy of Sciences,
Beijing 100049, China}

\begin{abstract}
An unconditionally secure authority-certified anonymous quantum key distribution scheme using conjugate coding is presented, base on which we construct a quantum election scheme without the help of entanglement state. We show that this election scheme ensures the completeness, soundness, privacy, eligibility, unreusability, fairness and verifiability of a large-scale election in which the administrator and counter are semi-honest. This election scheme can work even if there exist loss and errors in quantum channels. In addition, any irregularity in this scheme is sensible.
\end{abstract}


\end{frontmatter}

% \linenumbers

\section{Introduction}
Achieving security and privacy simultaneously is a problem that must be solved in a voting system if the system is to be used for serious large-scale government elections\cite{MI11}. In a traditional paper ballots method, the voting system often uses a locked box to solve this problem: After filling out the ballots, the voters put the ballots into the box one by one so as to guarantee the privacy of the voters. With the advent of information age, more and more daily life matters are transacted through digital communication network. Electronic voting scheme is appealing as an alternative to the traditional paper ballots method.

The first cryptographic voting protocol\cite{CD81} was suggested by Chaum in 1981. This protocol uses public key cryptography and rosters of digital pseudonyms to conceal the identity of votes, but it does not guarantee that the identity of voters cannot be traced. Later Chaum proposed a protocol that provided unconditionally security against tracing the voting\cite{CD88}. However, elections based on this protocol can be disrupted by a single voter who can break RSA. Paper\cite{ChD88} suggested an unconditionally secure solution of the so-called dinning cryptographers problem that can be used to guarantee unconditionally anonymity of the voters. Privacy in this scheme is assured, but the protocol cannot avoid a voter voting an arbitrary number times, so cheating is easy. Solving the privacy and fairness problems at the same time seems to be a drawback of cryptographic voting schemes. Over the years, much attention has been paid to electronic voting schemes\cite{FOO93,SA96,SK95,CC96,IKSA03,JKB11,KW11,MI11}. In practice, an electronic voting scheme can use blind signature, homomorphic-encryption, and other classical cryptographic techniques. For example, a secure electronic voting scheme based on blind signature was presented in \cite{IKSA03}, and paper\cite{FOO93} presented a practical secret voting scheme which is suitable for large scale election and solves the privacy and fairness problems at the same time. However, the security of these schemes depends on the solving of certain difficult mathematical problems, such as factoring large numbers and solving discrete logarithms. It could be easily broken with the emergence of quantum computers, thus allowing an attacker to disturb the election.

The physics of quantum cryptography pioneers a new way of the election. Different from the classical cryptography, the security of quantum cryptography is based on the fundamental laws of quantum physics to provide unconditional security\cite{BR03,CL05,DLZZ05,WW10}. For example, quantum key distribution can use the laws of physics as the basis for a scheme to distribute secure cryptographic keys\cite{GRTZ02}. Hence, more and more researchers show their interests in quantum election.

As in an election scheme it is more desirable that the identity of a voter, rather than his vote, be kept secret, quantum anonymous transmission\cite{CW05} seems to be an inspiration for quantum election. A few quantum election protocols in the context of this notion have been proposed recently: Vaccaro et al.\cite{VSC07} proposed quantum protocols for voting and surveying; a key feature of their schemes is the use of entangled states to ensure that the votes are anonymous and to allow the votes to be tallied. Two similar protocols\cite{DPT06,HZBB06} suggested the voters to apply local operations to an entangled state to guarantee the anonymity of participants in voting procedures, paper\cite{HZBB06} also presented a method of preventing voters from casting more than one vote. Another similar protocol based on quantum mechanics had been proposed in paper\cite{BBHZ11}, which used quantum mechanics to maintain the privacy of the voters. In 2011, Horoshko and Kilin\cite{HK11} proposed a new protocol for quantum anonymous voting, their protocol protected both the voters from a curious tallyman and all the participants from a dishonest voter in unconditional way. All the mentioned protocols provides unconditional anonymity of voting, but every voter in these protocols can only decide to vote for two alternatives, which may be impractical in an election that has many candidates. Unlike these schemes that using quantum entanglement for obtaining anonymity, paper\cite{LZ09} and paper\cite{OST08} presented quantum voting schemes that involving many candidates. In paper\cite{OST08} a blank ballot was an unknown quantum state for a voter, and the voter randomized his ballot to guarantee anonymity. But there was still a problem: as there was no way for a voter to verify whether his/her ballot had been counted correctly, the voter could not make sure that he/she had voted successfully.

In this paper we present a quantum election scheme based on anonymous quantum key distribution. Our scheme should be more easier to implement compared with the schemes in paper\cite{VSC07,DPT06,HZBB06,BBHZ11,HK11,LZ08,LZ09}, because we do not use quantum entanglement and the reading of the result do not need complicated measurement. In addition, the candidates in our scheme are not limited to two alternatives. Moreover, our scheme can ensure that each eligible voter can make sure that his/her ballot has been counted successfully compared with the scheme in Ref.\cite{OST08}.

The rest of this paper is organized as follows. In the next section, we present the unconditionally secure anonymous quantum key distribution. In section 3, we present our quantum election scheme in detail. In Section 4, we analyze the security of the scheme. Subsequently, we discuss the comparison between our scheme and the previous traditional and quantum election schemes in section 5. Finally, we make conclusions in section 6.

\section{Anonymous quantum key distribution}
\subsection{AQKD protocol}
To start with, we present a protocol for anonymous quantum key distribution with unconditional security.

We use the notation $\|$ to denote the concatenation of strings and $E_k[M]$ to denote an unconditionally secure symmetric encryption algorithm which encrypts message M by using one-time-use key k, and the length of k is equal to the length of m.

Suppose someone $U_i$ wants to anonymously establish a key with a certain entity Charlie with the help of another entity Bob. For this he uses the following protocol:
\newtheorem{protocol}{Protocol}
\begin{protocol}
{\bf AQKD($U_i$)}
\end{protocol}
Prerequisite: $U_i$ has established a 10m-bit key $P_{bi}$ with Bob by directly contacting or using an unconditionally secure quantum key distribution protocol\cite{BB84,E91,B92}. Where $P_{bi}=(p_i\|x_i\|y_i\|z_i)$. Here the entity Bob acts as the administrator, he only helps the users who pre-sharing $P_{b_*}$ with him to run the AQKD protocol, in other words, Bob will certify the authority of each user $U_i$ at the beginning of the protocol.

{\bf Step 1.} $U_i$ sends $p_i$ to Bob via a secure classical channel.

{\bf Step 2.} Bob checks whether $p_i^\prime$ is correct after he receiving it, if it is correct, Bob establishes a m-bit key $k_c$ and a (9+1)m-bit key $k_{bc}$ with Charlie by directly contacting or using an unconditionally secure quantum key distribution protocol. Then he computes
\begin{equation}
\begin{split}
X&=E_{k_{bc}}[k_c\|S_i]\\
   &=E_{k_{bc}}[k_c\|(x_i\|y_i\|z_i)]
\end{split}
\end{equation}
and sends it to Charlie via a secure classical channel.

{\bf Step 3.} When Charlie receives $X^\prime$, he uses $k_{bc}$ to decrypt it. After gaining the outcome $(k_c^\prime\|S_i^\prime)$, he checks whether $k_c^\prime$ is equal to $k_c$. If the check succeeds, Charlie can make sure that his outcome $S_i^\prime$ is equal to Bob's plaintext $S_i$.

{\bf Step 4.} $U_i$ randomly chooses two strings $r_{i1}, r_{i2}\in\{0,1\}^{3m}$ to generate his qubits $|\alpha_i\rangle$ by the method of conjugate coding\cite{WS83}:
\begin{equation}
|\alpha_i\rangle=H^{r_{i1}}|r_{i2}\rangle=\otimes_{j=1}^{3m} (H^{r_{i1}^j}|r_{i2}^j\rangle)
\end{equation}
where $r_{i1}^j$,$r_{i2}^j$ denote the j-th bit of $r_{i1}$,$r_{i2}$, and $H^{r_{i1}}=\otimes_{j=1}^{3m} H^{r_{i1}^j}$,
$H^0=I=\left(
                   \begin{array}{cc}
                     1 & 0 \\
                     0 & 1 \\
                   \end{array}
                 \right)
$, $H^1=H=\dfrac{1}{\sqrt{2}}\left(
                               \begin{array}{cc}
                                 1 & 1 \\
                                 1 & -1 \\
                               \end{array}
                             \right)
$.

Then $U_i$ anonymously transmits $|\alpha_i\rangle$ to Charlie via an error-free quantum channel.

{\bf Step 5.} Charlie randomly chooses the measuring basis to perform single-particle measurements the qubits $|\alpha_i\rangle^{\prime}$ that he received. After gaining the outcome $r_{i2}^\prime$, he randomly chooses a subset $\sigma_i$ of the set $\{1,2,\cdots,3m\}$, where $\sigma_i=\{d_i^1,d_i^2,\cdots,d_i^m\}$, with $1\leqslant d_i^1\textless d_i^2\textless\cdots\textless d_i^m\leqslant 3m$;
then generates the string $f_i\in\{0,1\}^{3m}$, where
\begin{equation}
f_i^j=\begin{cases}
0,\ \ \ {\rm for} \ j\in\sigma_i;\\
1,\ \ \ {\rm for} \ j\notin\sigma_i;\\
\end{cases}
\end{equation}
where $j\in\{1,2,\cdots,3m\}$ throughout. Then he computes
\begin{flalign}
F_i&=(r_{i2}^\prime)^{d_i^1}(r_{i2}^\prime)^{d_i^2}\dots(r_{i2}^\prime)^{d_i^m}
\end{flalign}

and publicly publishes the group $(F_i,f_i)$ with the corresponding measuring basis of the qubits $|\alpha_i\rangle^{\prime}$.

{\bf Step 6.} $U_i$ extracts $\sigma_i=\{d_i^1,d_i^2,\cdots,d_i^m\}$ through the string $f_i$ that published by Charlie and computes
\begin{flalign}
F_i^\prime&=(r_{i2})^{d_i^1}(r_{i2})^{d_i^2}\dots(r_{i2})^{d_i^m}
\end{flalign}

if $F_i^\prime$ satisfies the following conditions:
\begin{equation}
\text {if Charlie measures } |\alpha_i^{d_i^j}\rangle^{\prime}  \text {correctly, then } F_i^{\prime d_i^j}=F_i^{d_i^j} \\
\end{equation}
where $j\in\{1,2,\cdots,m\}$ throughout, $U_i$ generates the string $r_{i3}\in\{0,1\}^{2m}$, where
\begin{equation}
r_{i3}^{e_i^j}=\begin{cases}
1,\ \ \ \text {for Charlie measuring } |\alpha_i^{e_i^j}\rangle^{\prime} \text { correctly;}\\
0,\ \ \ \text {for Charlie measuring } |\alpha_i^{e_i^j}\rangle^{\prime} \text { incorrectly;}\\
\end{cases}
\end{equation}
where $j\in\{1,2,\cdots,2m\}$ throughout, with
\begin{equation*}
\{e_i^1,e_i^2,\cdots,e_i^{2m}\}=\{1,2,\cdots,3m\}-\sigma_i,\ \text {and}\  e_i^1\textless e_i^2\textless \cdots \textless e_i^{2m}.
\end{equation*}
Subsequently $U_i$ computes
\begin{equation}
Y_i=E_{z_i}[y_i\|r_{i3}]
\end{equation}
and anonymously sends $(x_i, Y_i)$ to Charlie via a secure classical channel.

{\bf Step 7.} When receives $(x_i^\prime,Y_i^\prime)$, Charlie checks whether he has accepted $x_i^\prime$ before. If he has not accepted $x_i^\prime$ before, he verifies whether $x_i^\prime$ is equal to $x_i$ that recorded in his database. If it is not, Charlie rejects to accept $Y_i^\prime$; otherwise Charlie uses $z_i$ to decrypt $Y_i^\prime$. With the outcome $(y_i^\prime\|r_{i3}^\prime)$, Charlie checks whether $y_i^\prime$ is equal to $y_i$. If it is true, Charlie uses $r_{i3}^\prime$ to extract the final key $G_i$, which made up by the measurement outcome of the qubits that Charlie has chosen the correct measuring basis.

Now the protocol is completed, $U_i$ establishes a key string $G_i$ with Charlie anonymously.
\subsection{Security analysis}
As the qubits $|\alpha_i\rangle$ in the protocol is an unknown quantum state to others, it is impossible to make a copy.
An attacker has to measure the initial qubits if he wants to get some useful information. However, the attacker does not know the encoding
basis of the qubits, he will choose the wrong measuring basis with the probability $50\%$ for each qubit, then introduce
no less than $25\%$ error rate in the string $F_i^\prime$ and to be detected by $U_i$ in step 6. The probability that others
except $U_i$ and $V_i$ get the correct x-bit key $G_i$ is not more than $(\frac{1}{2})^x$, when x is large enough, the probability is negligible.
So nobody except $U_i$ and Charlie knows the value of $G_i$. As Charlie is not aware of $U_i$'s identity $ID_i$, $U_i$ has established
a x-bit key $G_i$ with Charlie anonymously.

\subsection{Practicability analysis}
The protocol can also work even if there exist loss and errors in quantum channels. In order to establish an error-free key, Charlie can publish all the serial numbers and measuring basis of the qubits that he has received, then publish the check string $F_i$ and its serial numbers $f_i$ in step 5. If the error rate of $U_i's$ string $F_i^\prime$ is acceptable, $U_i$ can make sure that the key distribution is successful. After verifying that the key distribution is successful, $U_i$ extracts $r_{i3}$ and computes the key string $G_i$ which contains acceptable errors from $r_{i2}$ by using the serial numbers and measuring basis that published by Charlie, and subsequently uses $G_i$ and a key redistribution protocol\cite{YWL02} to establish the final error-free key $k_{ic}$:

(1) $U_i$ randomly chooses a string as the final key $K_{ic}$, then uses a classic error correction coding(ECC) to encode it. We use $D_i$ to denote the error correction code of $K_{ic}$.
Then $U_i$ computes
\begin{equation}
Y_i=E_{z_i}[y_i\|r_{i3}], \ \ \ Z_i=G_i\oplus D_i
\end{equation}
and sends $(x_i, Y_i,Z_i)$ to Charlie.

(2) When receives $(x_i^\prime,Y_i^\prime,Z_i^\prime)$, Charlie checks whether he has accepted $x_i^\prime$ before. If he has not accepted $x_i^\prime$ before, he verifies whether $x_i^\prime$ is equal to $x_i$ that recorded in his database. If it is not, Charlie rejects to accept $Y_i^\prime$; otherwise Charlie uses $z_i$ to decrypt $Y_i^\prime$. With the outcome $(y_i^\prime\|r_{i3}^\prime)$, Charlie checks whether $y_i^\prime$ is equal to $y_i$. If it is, Charlie uses $r_{i3}^\prime$ to extract $G_i^\prime$, then he computes
\begin{equation}
D_i^\prime=Z_i^\prime \oplus G_i^\prime=D_i\oplus e
\end{equation}
then Charlie decodes $D_i^\prime$ to get the final error-free key $K_{ic}$.

Obviously, if the error correction ability of the chosen error correction code is strong enough to correct the errors in $D_i^\prime$, Charlie will get the correct $k_{ic}$ at last.

Generally, a quantum key distribution(QKD) protocol is a key expansion protocol, in which the two parties in the protocol should pre-share a short key, which is used for identification and authentication of classical communication, to distribute a new longer key. In our anonymous quantum key distribution(AQKD), $U_i$ can anonymously establish a key with Charlie with the help of Bob under the condition that there is no key shared between $U_i$ and Charlie in advance. In addition, the final key between $U_i$ and Charlie is private to others(Bob included). The AQKD protocol plays a unique role in some cases(such as in an anonymous election scheme) although the final key that generated by the protocol is shorter than the keys that has been used up between $U_i$ and Bob, Bob and Charlie during the protocol.

\section{Quantum election scheme}

\subsection{Security definition and notations}
There are many properties that a perfect voting scheme should have. Reviewing the properties that described in the literatures\cite{CD88,FOO93,CC96}, we define a secure quantum election scheme should have the following properties:

$(1)$ Completeness: All valid votes are counted correctly.

$(2)$ Soundness: The dishonest voter cannot disrupt the voting.

$(3)$ Privacy: The content of the ballots is invisible to others.

$(4)$ Eligibility: Only eligible voters are permitted to vote.

$(5)$ Unreusability: Each eligible voter can vote successfully only once.

$(6)$ Fairness: Nothing must affect the voting.

$(7)$ Verifiability: Each eligible voter can check whether his/her ballot has been counted successfully.

We use ${ID}_i$ to represent the identity of the eligible voter $V_i$.

\subsection{Quantum election scheme}
A complete voting process in our election scheme includes four phases: initial phase, authentication phase, key distribution phase and voting phase. Several voters $V_j$, j=1,2,$\cdots$,N, the voting administrator Bob, and the counter Charlie are also involved.

In order to ensure the successful implement of the election, we assume the two parties Bob and Charlie in our scheme are semi-honest(honest but curious). In addition, they are two independent parties, that means they will not collaborate to trace a ballot.
\subsubsection{Initial phase}
The initial phase is as follows:

Before the authentication phase, the voting administrator Bob will publish a set $\mathcal{S}\subset \{0,1\}^s$. Each element of the set is a s-bit string that randomly chosen by Bob to represent an eligible candidate. In addition, Bob will announce the corresponding candidate of each element. In the election scheme, an eligible voter $V_j$ chooses the corresponding element of one candidate as his ballot $v_j$. We assume s is large enough to ensure the probability that a random s-bit string be an element of set $\mathcal{S}$ is negligible.

The voting administrator Bob establishes a key $k_{bj}$, where $k_{bj}=(k_j\|a_j\|b_j\|c_j)$, with every eligible voter $V_j$, j=1,2,$\cdots$,N, by directly contacting or using an unconditionally secure quantum key distribution protocol\cite{BB84,E91,B92}. All the four parts of $k_{bj}$ are selected uniquely for $V_j$.

All these work should be completed in advance.
\subsubsection{Authentication phase}
When the eligible voter $V_j$ wants to vote, he sends a request by sending the group $(ID_j,k_j)$ to Bob.

When Bob receives $({ID}_j,k_j)$ and the corresponding qubits, he checks whether $V_j$ has successfully applied for voting before. If $V_j$ has applied successfully before, Bob rejects to accept his request; otherwise, Bob verifies the string $k_j$, if it is correct, Bob accepts $V_j$'s request and records the corresponding $(i,ID_j,a_jb_jc_j)$ in a table with number i. (See Table.1)

At the end of the authentication phase, Bob announces the number of the verified voters(we denote it by n) and publishes a set that contains all the verified $ID_j$, $j\in\{1,2,\cdots,N\}$. Now the scheme turns to the key distribution phase.

\begin{center}
\tabcolsep=15pt
\small
\renewcommand\arraystretch{1.2}
\begin{minipage}{15.5cm}{
\small{\bf Table 1.} Bob's list of the verified voters.}
\end{minipage}
\vglue5pt
\begin{tabular}{| c | c | c |}
\hline %%
{Entry} & {Identity} &{Related information} \\
 \hline
  {1} & {$ID_t$} & {$S_1:(a_t,b_t,c_t)$}\\
 {$\vdots$}&{$\vdots$} &{$\vdots$}\\
 {i} & {$ID_j$} & {$S_i:(a_j,b_j,c_j)$}\\
 {$\vdots$}&{$\vdots$}&{$\vdots$} \\
 {n} & {$ID_l$} & {$S_n:(a_l,b_l,c_l)$}  \\
\hline
\end{tabular}
\end{center}
\subsubsection{Key distribution phase}
After dealing with all the applied voter $V_j$'s requests $(ID_j,k_j)$, $j\in\{1,2,\cdots,N\}$, Bob helps each verified voter $V_j$ to execute an anonymous quantum key distribution to establish a 2s-bit key $K_{ic}$ between $V_j$ and Charlie by running
\begin{equation*}
{{\bf AQKD}(U_i)}
\end{equation*}
with $U_i=V_j$.

After establishing key strings with all the verified voters, Charlie publicly announces that the key distribution phase is completed, the scheme turns to the voting phase.

\begin{center}
\tabcolsep=15pt
\small
\renewcommand\arraystretch{1.2}
\begin{minipage}{15.5cm}{
\small{\bf Table 2.} List of Charlie's key strings.}
\end{minipage}
\vglue5pt
\begin{tabular}{| c | c | c |}
\hline %%
  \multicolumn{2}{|c|}{($K_{*c}^\prime$)} & \multirow{2}*{Remark($a_*$)}\\
\cline{1-2}
{$K_{*cL}^\prime$} & {$K_{*cR}^\prime$}& \\
 \hline
 {$K_{1cL}^\prime$} & {$K_{1cR}^\prime$} &{$a_t$}\\
 {$\vdots$} & {$\vdots$}&{$\vdots$}\\
 {$K_{icL}^\prime$} & {$K_{icR}^\prime$} &{$a_j$}\\
 {$\vdots$} & {$\vdots$}& {$\vdots$}\\
 {$K_{ncL}^\prime$} & {$K_{ncR}^\prime$} &{$a_l$}\\
\hline
\end{tabular}
\end{center}

Here $K_{*c}^\prime=K_{*cL}^\prime\|K_{*cR}^\prime$, and $K_{*cL}^\prime,K_{*cR}^\prime \in \{0,1\}^s$.

\subsubsection{Voting phase}
When the verified voter $V_j$ makes sure that he has anonymously established a 2s-bit key $k_{ic}$ with Charlie successfully, he chooses an element from set $\mathcal{S}$ as his ballot $v_j$ and anonymously sends $(K_{icL},E_{K_{icR}}[v_j])$ to Charlie.

After receiving $(K_{icL},E_{K_{icR}}[v_j])$, Charlie checks whether $K_{icL}$ is in the first column of Table.2. If it is, Charlie checks whether $K_{icL}$ has been accepted before. If not, Charlie extracts the corresponding group $(K_{icL}^\prime, K_{icR}^\prime,a_j)$, where $K_{icL}^\prime=K_{icL}$, from Table.2 and uses $K_{icR}^\prime$ to decrypt $E_{K_{icR}}[v_j]$. If the outcome $v_j\in \mathcal{S}$, Charlie counts $v_j$ and accepts $K_{icL}$.

After dealing with all the groups $(K_{icL},E_{K_{icR}}[v_j])$, Charlie deletes the lines in which the first item $K_{*cL}$ has not ever been accepted by him from Table.2.  Then Charlie publishes a set which contains all the corresponding strings $\{a_j\}$ of the accepted $K_{icL}^\prime$. Both Bob and the verified voter $V_j$ can verify whether $V_j$ has voted successfully by searching the value of $a_j$ in this set. Bob communicates with each verified voter $V_i$ that has not voted successfully to establish a new $k_{bi}$, then allows these failed voters for a new round of election.

While all the verified voters vote successfully, Charlie counts the number of each candidate's ballots. Subsequently, he randomly arranges all the accepted groups $(K_{icL},v_j)$ and publicly publishes them. The scheme is now completed.(See Table.3)

\begin{center}
\tabcolsep=15pt
\small
\renewcommand\arraystretch{1.2}
\begin{minipage}{15.5cm}{
\small{\bf Table 3.} List of Ballots (for public).}
\end{minipage}
\vglue5pt
\begin{tabular}{| c | c |}
\hline %%
  {Ballot}& {Additional information}\\
 \hline
 {$\vdots$}& {$\vdots$}\\
 {$v_j$} & {$K_{icL}$} \\
 {$\vdots$}& {$\vdots$}\\
\hline
\end{tabular}
\end{center}
\section{Security analysis of the election scheme}
In this section, we are to discuss the security of the quantum election scheme. We will prove that the scheme has the properties of completeness, soundness, privacy, eligibility, unreusability, fairness and verifiability when the administrator and the counter are semi-honest.

\subsection{Completeness}
The completeness means that all the valid ballots in the scheme are counted correctly. In this quantum election scheme, if all the three participants(the voters, the voting administrator, and the counter) are honest, after receiving an eligible voter $V_j$'s voting request, Bob will send the messages $S_i$ to Charlie honestly in the key distribution phase; when $V_j$ makes sure that the anonymous key distribution protocol is successful, he will honestly send the messages $(K_{icL},E_{K_{icR}}[v_j])$ to Charlie; After extracting $V_j$'s ballot $v_j$ from the cipher $E_{K_{icR}}[v_j]$, Charlie will count it honestly. So the result of the election is trustable.
\subsection{Soundness}
While a dishonest voter $V_j$ wants to disrupt the voting, he may keep not sending the valid group $(K_{icL},E_{K_{icR}}[v_j])$ or sending an invalid group $(K_{icL},E_{K_{icR}}[v_x])$, where $v_x$ represents an invalid candidate, to Charlie in the voting phase, then applying for a new round of voting. In the first case, $V_j$ will be detected by Bob, because before a new round of voting Bob has to establish a new key with $V_j$. If $V_j$ fails too much times, Bob will detect him/her. In the second case, $V_j$ will be detected by Charlie in the voting phase. Therefore, the dishonest voter cannot disrupt the election.
\subsection{Privacy}
In the key distribution phase, the voter $V_j$ sends his quantum blank vote $|\eta_j\rangle$ to the voting administrator Bob. As $|\eta_j\rangle$ is independent of $v_j$ before the voting phase, neither an attacker nor a curious Bob can get anything about $v_j$ even if they get the entire state $|\eta_j\rangle$. As the anonymous quantum key distribution protocol in this scheme is unconditionally secure, only Charlie and $V_j$ know the correct $K_{ic}$ if $V_j$ ensures that his anonymous quantum key distribution is successful; although Charlie knows the value of $v_j$, he cannot get the corresponding $ID_j$. Hence, nobody except $V_j$ can match $ID_j$ with the ballot $v_j$.

In this scheme it is possible to assume that Bob and Charlie are independent parties,i.e., they will not cooperate to trace the ballots.
\subsection{Eligibility}
The eligibility of an voting scheme means only eligible voters are permitted to vote, and it is impossible for an ineligible voter to forge a valid ballot. As described in the authentication phase, Bob will check the identity of a voter and reject to accept his/her voting request if the voter cannot pass through the identity authentication, therefore only eligible voters are permitted to vote.

If an ineligible voter $V_e$ wants to vote successfully, he has only two ways, one is to personate an eligible voter $V_j$ in the voting phase, the other is to generate a valid ballot in the voting phase. In the first case, suppose $V_e$ has intercepted $V_j$'s qubits $|\eta_i\rangle$, he replaces $|\eta_i\rangle$ by $|\eta_e\rangle$ and sends $|\eta_e\rangle$ to Charlie. In order to vote successfully, $V_e$ has to computes
\begin{equation*}
Y_e=E_{c_j}[b_j\|r_{e3}]
\end{equation*}
and send the group $(a_j, Y_e)$ to Charlie in the key distribution phase. However, $V_e$ has no way to get the value of $(a_j,b_j,c_j)$, thus it is impossible for him to compute the correct $Y_e$, so he cannot personate $V_j$ successfully; in the second case, $V_e$ may monitor the classical channel to get a verified voter $V_j$'s group $(K_{icL},E_{K_{icR}}[v_j])$, and send his group $(K_{icL},E_{K_{icR}}[v_e])$ to Charlie, if his group reaches earlier than $V_j$'s group, Charlie will accept $(K_{icL},E_{K_{icR}}[v_e])$. However, it is impossible for $V_e$ to compute $E_{K_{icR}}[v_e]$, for he does not has the encryption key $K_{icR}$. At the same time, as s is large enough, the probability that a random s-bit string $v_*$ be a valid ballot is negligible. Therefore it impossible for an attacker to forge a valid vote.

In this paper we assume the administrator Bob and the counter Charlie are semi-honest. Actually, a malicious administrator Bob or a malicious counter Charlie may also try to forge valid votes: as Bob knows $(a_j,b_j,c_j)$, he can easily personate the eligible voter $V_j$ and vote successfully; in addition, a malicious Charlie may tamper the ballot of an eligible voter. Although these irregularities will be discovered by $V_j$ at the end of the scheme, there is no way to distinguish between a dishonest voter and a malicious Bob or a malicious Charlie. To avoid this, we can use a distributed scheme, in which several independent parties, e.g., several candidates of the election, collaborate to act as the voting administrator and another several independent parties collaborate to act as Charlie. They won't cooperate up forging invalid votes or tracing valid ballots. The distributed scheme ensures the credibility of the voting administrator and the counter.
\subsection{Unreusability}
In the authentication phase, before responding to an eligible voter $V_j$'s voting request, Bob will check whether $V_j$ has successfully applied for voting before to avoid accepting $V_j$'s another voting request; In addition, before accepts a group $(K_{icL},E_{K_{icR}}[v_j])$, Charlie will check whether the corresponding $K_{icL}$ has been accepted before to avoid accepting the group twice. Moreover, before a new round of voting, Charlie will delete the lines in which the first item has not ever been accepted from Table.2 to avoid accepting an overdue group $(K_{icL},E_{K_{icR}}[v_j])$ that had not been send by $V_i$ in the former round of voting. Hence, each eligible voter can vote successfully only once.
\subsection{Fairness}
The fairness means that nothing must affect the voting, especially the counting of ballots doesn't affect the voting. In this scheme, the voting phase is done after the authentication phase and key distribution phase, and Charlie will not disclose the intermediate result of the election to others before the whole scheme is completed. Therefore the previous voters will not affect the subsequent voters. So this scheme is fair for all voters.
\subsection{Verifiability}
At the end of the scheme, Charlie will publish Table.3 publicly to all the voters. Each eligible voter $V_j$ can search his group $(K_{icL},v_j)$ in this table. If $(K_{icL},v_j)$ is in the table, $V_j$ can make sure that his vote has been counted correctly.
\section{Discussion}
%Unlike traditional electronic voting schemes that based on classical cryptography, the security of our scheme is guaranteed by the fundamental laws of quantum mechanics, the quantum blank votes in the scheme are unknown qubits to others, so it is impossible to make a copy of it. In a traditional electronic voting scheme, there is no reliable way of knowing whether someone has successfully eavesdropped in the channel, so an attacker can monitor the channel to get the all the information that communicated in the scheme. In our scheme, quantum states are used for unconditionally secure classical information transmission. As the quantum blank vote that transformed in the quantum channel are independent of the voter's choice, an attacker cannot get any useful information even if he intercepts the entire states. In addition, as the attacker does not know the encoding basis of the qubits, it is impossible for him to perform a correct measurement; moreover, his incorrect measurement will change the initial quantum blank vote and then to be detected by the voter in the verifying phase. Once an eligible voter detects that his quantum blank vote has been changed, he will not send the encrypted ballot to the counter. As a result, the attacker cannot get any available information.

In a traditional electronic voting scheme, there is no reliable way of knowing whether someone has successfully eavesdropped in the channel. Unlike the traditional electronic voting schemes that based on classical cryptography, any eavesdropping behavior of quantum information in our scheme will be discovered. As the quantum blank votes in the scheme are unknown qubits to others, it is impossible to make a copy of them, so the attacker has to measure the initial qubits if he wants to get some useful information. However, the attacker does not know the encoding basis of the qubits, it is impossible for him to perform a correct measurement; moreover, his incorrect measurement will change the initial quantum blank vote and then to be detected by the voter in the key distribution phase. Once an eligible voter detects that his quantum blank vote has been changed, he will not send the encrypted ballot to the counter. In addition, as the quantum blank vote that transformed in the quantum channel are independent of the voter's choice, an attacker cannot get any useful information even if he intercepts the entire states. As a result, the attacker cannot get any available information.

Compared with the previous quantum voting schemes in paper\cite{VSC07,DPT06,HZBB06,BBHZ11,HK11,LZ08,LZ09}, Our scheme should be more easier to implement because we do not quantum entanglement. In view of the problem of verifiability that not mentioned in Ref.\cite{OST08}, we also discuss the verifiability of our scheme to make sure that every voter can check whether his/her vote has been counted successfully.

In this paper we present an unconditionally secure quantum election scheme. As there is no reliable way to prevent the participants from copying the classical messages, we assume the administrator and counter in the scheme is semi-honest, we can use a distributed scheme to ensure the credibility of the voting administrator and the counter. A malicious voter cannot disrupt the voting and any irregularity in the scheme is sensible. The scheme can work even if the quantum channels exist certain loss and the error rate.
\section{Conclusion}
We present an unconditionally secure quantum election scheme based on anonymous quantum key distribution. Our scheme ensures the completeness, the soundness, the privacy, the eligibility£¬the unreusability, the fairness and the verifiability of the quantum election scheme in which the administrator and counter are semi-honest. A malicious voter cannot disrupt the voting and any irregularity in the scheme is sensible. The key technique of our scheme is using unknown quantum states to execute authority-certified anonymous quantum key distribution with unconditional security. Different from previous quantum voting schemes, we do not use quantum entanglement, and each eligible voter can check whether his/her vote has been counted successfully.

\section*{Acknowledgement}
This work was supported by the National Natural Science Foundation of China under Grant No.61173157.


\begin{thebibliography}{}
%\softraggedright
\itemsep=-4pt plus.2pt minus.2pt  %% sets the vertical space between items
\small
\bibitem{WS83}Wiesner S 1983 {\textit Sigact News} {\bf 15}
78
\bibitem{BB84}Bennett C H and Brassard G 1984 {\textit Advances in Proceedings of the IEEE International Conference on Computers, Systems and Signal Processing}(New York:IEEE) p.175
\bibitem{E91}Ekert A K 1991 {\textit Phys. Rev.} Lett.  {\bf 67}
661
\bibitem{B92}Bennett C H 1992 {\textit Phys. Rev.} Lett.  {\bf 68}
3121
\bibitem{GRTZ02}Gisin N, Ribordy G, Tittel W, and Zbinden H 2002 {\textit Rev. Mod. Phys.} {\bf 74}
145
\bibitem{CW05}Christandl M and Wehner S 2005 {\textit LNCS} {\bf 3788}
217
\bibitem{BR03}Boykin P O and Roychowdhury V 2003 {\textit Phys. Rev.} A {\bf 67}
042317
\bibitem{CL05}Chen K and Lo H K 2005 arXiv:0404133 [quant-ph]
\bibitem{DLZZ05}Deng F G, Li X H, Zhou H Y, and Zhang Z J 2005 {\textit Phys. Rev.} A {\bf 72}
044302
\bibitem{WW10}Wang T Y and Wen Q Y 2010 {\textit Chin. Phys.} B {\bf 19}
060307
\bibitem{YWL02}Yang L, Wu L A and Liu S H 2002 {\textit Acta Physica Sinica} {\bf 51}
961
\bibitem{CD81}Chaum D 1981 {\textit Communications of the ACM} {\bf 24}
84
\bibitem{CD88}Chaum D 1988 {\textit Advances in Cryptology-EUROCRYPT'88, Lecture Notes in Computer Science} {\bf 330}  (Berlin: Springer-Verlag) p.177
\bibitem{ChD88}Chaum D 1988 {\textit Journal of cryptology} {\bf 1}
65
\bibitem{FOO93}Fujioka A, Okamoto T and Ohta K 1993 {\textit Lecture Notes in Computer Science} {\bf 718}
244
\bibitem{SA96}Salomaa A 1996 {Public-Key Cryptography} (2nd ed) (Berlin: Springer-Verlag) pp200--202
\bibitem{SK95}Sako K and Killian J 1995 {\textit Advances in Cryptology-CRYPTO'94} (Berlin: Springer-Verlag) p.411
\bibitem{CC96}Cranor L F and Cytron R K 1996 Washington University Computer Science Technical Report 1996
\bibitem{IKSA03}Ibrahim S, Kamat M, Salleh M, and Aziz S R A 2003 {\textit National Conference on Telecommunication Technology Proceedings} p.193
\bibitem{JKB11}Jafari S, Karimpour J, and Bagheri N 2011 {\textit International Journal on Computer Science and Engineering} {\bf 3}
2191
\bibitem{KW11}Kumar S and Walia E 2011 {\textit International Journal on Computer Science and Engineering} {\bf 3}
1825
\bibitem{MI11}Marius I and Ionut P 2011 {\textit Computer Science Master Research} {\bf 1}
67
\bibitem{VSC07}Vaccaro J A, Spring J, and Chefles A 2007 {\textit Phys. Rev.} A  {\bf 75}
012333
\bibitem{DPT06}Dolev S, Pitowsky I, and Tamir B 2006 arXiv:0602087 [quant-ph]
\bibitem{HZBB06}Hillery M, Ziman M, Buzek v, and Bielikov M 2006 {\textit Phys. Lett.} A {\bf 349}
75
\bibitem{BBHZ11}Bonanome M, Buzek V, Hillery M, and Ziman M 2011 {\textit Phys. Rev.} A {\bf 84}
022331
\bibitem{HK11}Horoshko D and Kilin S 2011 {\textit Phys. Lett.} A {\bf 375}
1172
\bibitem{LZ08}Li Y and Zeng G 2008 {\textit Optical Review} {\bf 15}
219
\bibitem{LZ09}Li Y and Zeng G 2009 {\textit Chinese Optics Letters} {\bf 7}
152
\bibitem{OST08}Okamoto T, Suzuki K, and Tokunaga Y 2008 NTT {\textit Thchnical Review} {\bf 6}

\end{thebibliography}
\end{document}